\documentclass[a4paper,11pt]{article}
\usepackage{jheppub} 
\usepackage{lineno}
\usepackage{natbib}
\usepackage{ulem}
\usepackage[T1]{fontenc}
\usepackage{xcolor}
\usepackage{slashed}
\usepackage{feynmp-auto}

\usepackage{subcaption}

\newcommand\aj{\ref@jnl{AJ}}
\newcommand\psj{\ref@jnl{PSJ}}
\newcommand\araa{\ref@jnl{ARA\&A}}
\newcommand\apj{\ref@jnl{ApJ}}
\newcommand\apjl{\ref@jnl{ApJL}} 
\newcommand\apjs{\ref@jnl{ApJS}}
\newcommand\ao{\ref@jnl{ApOpt}}
\newcommand\apss{\ref@jnl{Ap\&SS}}
\newcommand\aap{\ref@jnl{A\&A}}
\newcommand\aapr{\ref@jnl{A\&A~Rv}}
\newcommand\aaps{\ref@jnl{A\&AS}}
\newcommand\azh{\ref@jnl{AZh}}
\newcommand\baas{\ref@jnl{BAAS}}
\newcommand\icarus{\ref@jnl{Icarus}}
\newcommand\jaavso{\ref@jnl{JAAVSO}} 
\newcommand\jrasc{\ref@jnl{JRASC}}
\newcommand\memras{\ref@jnl{MmRAS}}
\newcommand\mnras{\ref@jnl{MNRAS}}
\newcommand\pra{\ref@jnl{PhRvA}}
\newcommand\prb{\ref@jnl{PhRvB}}
\newcommand\prc{\ref@jnl{PhRvC}}
\newcommand\prd{\ref@jnl{PhRvD}}
\newcommand\pre{\ref@jnl{PhRvE}}
\newcommand\prl{\ref@jnl{PhRvL}}
\newcommand\pasp{\ref@jnl{PASP}}
\newcommand\pasj{\ref@jnl{PASJ}}
\newcommand\qjras{\ref@jnl{QJRAS}}
\newcommand\skytel{\ref@jnl{S\&T}}
\newcommand\solphys{\ref@jnl{SoPh}}
\newcommand\sovast{\ref@jnl{Soviet~Ast.}}
\newcommand\ssr{\ref@jnl{SSRv}}
\newcommand\zap{\ref@jnl{ZA}}
\newcommand\nat{\ref@jnl{Nature}}
\newcommand\iaucirc{\ref@jnl{IAUC}}
\newcommand\aplett{\ref@jnl{Astrophys.~Lett.}}
\newcommand\apspr{\ref@jnl{Astrophys.~Space~Phys.~Res.}}
\newcommand\bain{\ref@jnl{BAN}}
\newcommand\fcp{\ref@jnl{FCPh}}
\newcommand\gca{\ref@jnl{GeoCoA}}
\newcommand\grl{\ref@jnl{Geophys.~Res.~Lett.}}
\newcommand\jcp{\ref@jnl{JChPh}}
\newcommand\jgr{\ref@jnl{J.~Geophys.~Res.}}
\newcommand\jqsrt{\ref@jnl{JQSRT}}
\newcommand\memsai{\ref@jnl{MmSAI}}
\newcommand\nphysa{\ref@jnl{NuPhA}}
\newcommand\physrep{\ref@jnl{PhR}}
\newcommand\physscr{\ref@jnl{PhyS}}
\newcommand\planss{\ref@jnl{Planet.~Space~Sci.}}
\newcommand\procspie{\ref@jnl{Proc.~SPIE}}

\newcommand\actaa{\ref@jnl{AcA}}
\newcommand\caa{\ref@jnl{ChA\&A}}
\newcommand\cjaa{\ref@jnl{ChJA\&A}}
\newcommand\jcap{\ref@jnl{JCAP}}
\newcommand\na{\ref@jnl{NewA}}
\newcommand\nar{\ref@jnl{NewAR}}
\newcommand\pasa{\ref@jnl{PASA}}
\newcommand\rmxaa{\ref@jnl{RMxAA}}

\title{\boldmath Ultra-Strongly Self-Interacting Dark Matter: From Phenomenology to Astrophysical Observables}

\author[1,2]{M. Grant Roberts,}
\author[1,2]{Wolfgang Altmannshofer,}
\author[1,2]{Pierce Giffin,}
\author[1,2]{Stefano Profumo}

\affiliation[1]{Department of Physics, 1156 High St., University of California Santa Cruz, Santa Cruz, CA 95064, USA}

 \affiliation[2]{Santa Cruz Institute for Particle Physics, 1156 High St., Santa Cruz, CA 95064, USA}
 
\emailAdd{migrober@ucsc.edu}
\emailAdd{waltmann@ucsc.edu}
\emailAdd{pgiffin@ucsc.edu}
\emailAdd{profumo@ucsc.edu}

\abstract{We develop a minimal, testable framework for two-component self-interacting dark matter (SIDM) in which a dominant, moderately self-interacting species coexists with an ultra-strongly self-interacting subcomponent (uSIDM). A light vector mediator induces velocity-dependent self-scattering, while early-universe dynamics - standard $2 \to 2$ annihilation supplemented by interconversion $\chi_1\chi_1 \to \chi_2\chi_2$ - determine the relic abundance analytically. From observations of dwarf and low surface brightness galaxy rotation curves, as well as strong cluster lensing, we place constraints on the microphysics parameters. From these constrained regions, we map the microphysics to effective \texttt{ETHOS} parameters and evolve the linear power spectrum in \texttt{CLASS}. We identify a region where: (1) the SIDM dominant component attains $\sigma_{\rm{eff}}/m = 20\textup{--}40~\cmg$ at dwarf velocities while satisfying cluster upper bounds $\sigma_{\rm{eff}}/m < 0.13~\cmg$; (2) a subpercent uSIDM fraction drives accelerated gravothermal collapse in early halos, providing seeds relevant to high-redshift quasar formation and ``little red dots''; and (3) the small-scale cutoff in the matter power spectrum remains consistent with Lyman-$\alpha$ and satellite counts, but exhibits non-standard features, potentially discernible with future observations. The allowed space can be organized by the mediator-to-DM mass ratio and the late-time uSIDM fraction, with a narrow window singled out by the combined cosmological and astrophysical requirements.}

\begin{document}

\newcommand{\mBH}{m_{\text{BH}}}
\newcommand{\mBHseed}{\mBH^{\text{seed}}}
\newcommand{\mBHobs}{\mBH^{\text{obs}}}
\newcommand{\mBHobsi}{m_{\text{BH},i}^{\text{obs}}}
\newcommand{\mBHtheory}{\mBH^{\text{theory}}}
\newcommand{\zcoll}{z_{\text{coll}}}
\newcommand{\zvir}{z_{\text{vir}}}
\newcommand{\zobs}{z_{\text{obs}}}
\newcommand{\cross}{\sigma/m}
\newcommand{\msun}{M_{\odot}}
\newcommand{\tsal}{t_{\text{sal}}}
\newcommand{\trel}{t_{\text{rel}}}
\newcommand{\rhocrit}{\rho_{\text{crit}}}
\newcommand{\cmg}{\text{cm}^{2}\text{g}^{-1}}
\newcommand{\kms}{\text{km}~\text{s}^{-1}}
\newcommand{\angstrom}{\r{A}}
\newcommand\sbullet[1][.5]{\mathbin{\vcenter{\hbox{\scalebox{#1}{$\bullet$}}}}}
\newcommand{\chisq}{\chi^{2}}
\newcommand{\vmax}{V_{\rm{max}}}

\newcommand{\M}{\mathcal{M}}
\newcommand{\lag}{\mathcal{L}}
\newcommand{\lagdark}{\lag_{\rm{dark}}}

\newcommand{\lrb}[1]{\left[{#1}\right]}
\newcommand{\lrp}[1]{\left({#1}\right)}
\newcommand{\lrcb}[1]{\left\{{#1}\right\}}
\newcommand{\lrv}[1]{\left|{#1}\right|}
\newcommand{\lra}{\longrightarrow}

\newcommand{\mchi}{m_{\chi}}
\newcommand{\mchione}{m_{\chi_1}}
\newcommand{\mchitwo}{m_{\chi_2}}
\newcommand{\mphi}{m_{\phi}}
\newcommand{\alphachi}{\alpha_{\chi}}
\newcommand{\alphachione}{\alpha_{\chi_{1}}}
\newcommand{\alphachitwo}{\alpha_{\chi_{2}}}
\newcommand{\gchione}{g_{\chi_{1}}}
\newcommand{\gchitwo}{g_{\chi_{2}}}
\newcommand{\ma}{m_{A'}}
\newcommand{\Tchi}{T_{\chi}}
\newcommand{\mpl}{M_{\text{pl}}}

\newcommand{\sigmam}{\sigma/m}
\newcommand{\sigmaeff}{\sigma_{\text{eff}}}
\newcommand{\sigmavel}{\sigma_\text{1D}}
\newcommand{\sigmaeffm}{\sigmaeff/m}
\newcommand{\sigmaeffmchitwo}{\sigmaeff/\mchitwo}

\newcommand{\chibar}{\overline{\chi}}
\newcommand{\OmegaDM}{\Omega_{\text{DM}}}
\newcommand{\OmegaCDM}{\Omega_{\text{CDM}}}
\newcommand{\OmegaDMh}{\OmegaDM\text{h}^{2}}
\newcommand{\OmegaCDMh}{\OmegaCDM\text{h}^{2}}

\newcommand{\alphaethos}{\alpha_{\rm{ETHOS}}}
\newcommand{\vrel}{v_{\rm{rel}}}
\newcommand{\kinmix}{\varepsilon_{\gamma}}
\newcommand{\fend}{f_{\rm uSIDM}^{\rm \infty}}

\newcommand{\spr}[1]{{\color{red}\bf[SP: {#1}]}}
\newcommand{\grantcomment}[1]{{\color{blue}\bf[GR: {#1}]}}
\newcommand{\PG}[1]{{\color{purple}\bf[PG: {#1}]}}

\newcommand{\WA}[1]{{\color{orange}\bf[WA: {#1}]}}

\newcommand{\updatetext}[1]{{\color{blue}[{#1}]}}

\maketitle
\flushbottom

\section{Introduction}

The discovery of compact, red, kpc- to sub-kpc--scale sources at $z\gtrsim4$, commonly referred to as ``Little Red Dots'' (LRDs), has sharpened long-standing questions about the rapid assembly of supermassive black holes (SMBHs) in the early Universe and the physical drivers of small-scale structure in the dark sector~\cite{Kokorev_2024,Kocevski:2024zzz,Casey:2024olr}. One intriguing possibility is that a small subcomponent of the dark matter (DM) possesses ultra-strong self-interactions, triggering rapid gravothermal evolution and early core-collapse in nascent halos, thereby seeding massive black holes that can power LRD-like systems~\cite{Pollack:2014rja,Roberts_uSIDM,Roberts_LRD}. In this ultra-strongly self-interacting DM (uSIDM) picture, only a subpercent fraction of the DM needs very large self-scattering to core-collapse quickly; the remainder can follow the successful large-scale $\Lambda$CDM phenomenology \cite{Salucci:2018hqu}.

Independently, a broad body of work has shown that self-interacting DM (SIDM) with velocity-dependent cross-sections offers a coherent explanation for several small-scale tensions, ranging from core formation and the diversity of rotation curves in dwarfs and spirals to cluster-scale density profiles, without deviating from $\Lambda$CDM on large scales~\cite{Tulin:2017ara,Adhikari:2022sbh,2025PhRvD.112h3011Y,Kaplinghat:2015aga}. Observationally, dwarf and low-surface-brightness galaxies often exhibit slowly rising rotation curves, a phenomenology emphasized in early systematic studies of spiral galaxies~\cite{Persic:1995ru}. Recent analyses explicitly demonstrate how gravothermal evolution in SIDM can amplify halo-to-halo diversity and reconcile high-density outliers, while preserving good fits to galaxies with slowly rising rotation curves~\cite{Roberts_rotation_curve}. On the opposite end of the mass scale, strong-lensing studies of galaxy clusters and galaxies provide complementary bounds on SIDM at higher characteristic velocities, tightening the viable parameter space and informing the velocity dependence of the cross-section~\cite{Andrade:2020lqq,ODonnell_strong_lensing,Sagunski:2020spe}.

These astrophysical observables interface naturally with the gravothermal evolution of SIDM halos. For a wide class of velocity-dependent models, the late-time evolution - including the onset and timing of core collapse - exhibits approximate universality once cast in terms of an appropriate scattering timescale, enabling robust predictions for collapse redshifts at fixed halo assembly histories~\cite{Outmezguine:2022bhq,GadNasr2023}. In this setting, an ultra-strong subcomponent can undergo accelerated collapse even if it constitutes only a small fraction of the total DM, provided its cross-section per mass is sufficiently large at the relevant internal velocities. The resulting black-hole seeds form early (e.g., $z\sim10$), after which baryonic accretion and feedback can generate luminous, compact systems with LRD-like spectral and morphological properties, including heavy obscuration scenarios consistent with the absence of bright far-IR/sub-millimeter detections to date~\cite{Casey:2024olr,Roberts_LRD}.

In this work we develop and test a minimal, predictive framework in which a dominant, moderately self-interacting species coexists with a subpercent uSIDM subcomponent. We: (1) lay out the microphysical setup and relic-density accounting for a light-mediator model that simultaneously fixes the late-time uSIDM fraction and the velocity dependence of self-scattering; (2) connect to halo phenomenology across dwarf-to-cluster scales via rotation curves and strong-lensing constraints; and (3) delineate the combined viable parameter space, emphasizing the mediator-to-DM mass ratio and the late-time uSIDM fraction as organizing variables. We also show how future observational measurements of the linear matter power spectrum can probe this two-component uSIDM scenario if the dark photon decays to massless dark fermion states.


\section{Model Overview}\label{sec:model-overview}

We introduce our dark sector Lagrangian as follows: the fermion $\chi_{1}$ represents the uSIDM component while the fermion $\chi_{2}$ corresponds to SIDM. We assume a mass splitting where $\mchione > \mchitwo$, in order to suppress the $\bar\chi_2 \chi_2 \to \bar\chi_1 \chi_1$ conversion channel during freeze-out, see the discussion in Section~\ref{subsec:boltzmann-equations} below. We also introduce a dark photon $A'$, with mass $\ma$, to mediate the self-interactions of the dark matter particles:

\begin{eqnarray}
 &&\lagdark = \overline{\chi}_{1}\lrp{i\slashed{\partial} - \mchione}\chi_{1} + \overline{\chi}_{2}\lrp{i\slashed{\partial} - \mchitwo}\chi_{2} - \gchione\chibar_{1}A'^{\mu}\gamma_{\mu}\chi_{1} - \gchitwo\chibar_{2}A'^{\mu}\gamma_{\mu}\chi_{2}\label{eq:model-lagrangian}
\\
 && ~~~~~~~~~~~ -\frac{1}{4}F'^{\mu\nu}F'_{\mu\nu} - \frac{1}{2}m_{A'}^{2}A'^{\mu}A'_{\mu}\nonumber.
\end{eqnarray}

Due to the self interactions, one could imagine that the dark matter particles form bound states. We address this possibility in Appendix~\ref{appendix:bound-state-estimate}, and we find that our model will not form bound states. In addition to the interactions in Eq.~\eqref{eq:model-lagrangian}, we also include an interaction channel between the dark photon and a massless dark fermion $\psi$, $-g_{\psi}\bar{\psi}A'^{\mu}\gamma_{\mu}\psi$, which we explore in Section~\ref{subsec:matter-power-spectrum}. Phenomenologically, uSIDM is characterized by two key parameters: the uSIDM self interaction cross-section, $\sigmam$, and the uSIDM fraction today, $\fend$. The fraction $\fend$ quantifies the fraction of dark matter that is ultra-self-interacting, $\fend = \frac{\rho_{\chi_1}}{\rho_{\chi_1} + \rho_{\chi_2}} \leq 1$ (but typically $\fend \ll 1$) where $\rho_{\chi_i}$ is the mass density of each species. Motivated by the $f\cdot(\sigmam)$ scaling of the self-scattering optical depth and relaxation rate of a subdominant uSIDM component \cite{Pollack:2014rja}, we take the couplings of the two dark matter species to be related by $\gchitwo = \gchione f^{1/4}$, where $f \ll 1$, and in Section~\ref{sec:thermally-averaged-cross-sections}, we will show that $f$ is closely related to $\fend$. Thus, we obtain a strong hierarchy between the corresponding self-interaction cross-sections,

\begin{equation}
 \lrp{\sigmam}_{\rm{SIDM}} = \left(\frac{g_{\chi_2}}{g_{\chi_1}}\right)^4 \cdot\lrp{\sigmam}_{\rm{uSIDM}} = f\cdot\lrp{\sigmam}_{\rm{uSIDM}} ~,
\label{eq:cross-section-conversion}
\end{equation}
such that $\lrp{\sigmam}_{\rm{uSIDM}} \gg \lrp{\sigmam}_{\rm{SIDM}}$.

To get an initial insight, some SIDM models that interact via a Yukawa potential~\cite{Tulin:2013teo,Lankester--Broche:2025rkm} exist in the Born region of their respective parameter spaces; thus, we can get a rough idea of the non-relativistic differential scattering cross-section by considering the perturbative limit of the self-interaction for either $\chi_{1}$ or $\chi_{2}$. In the following equations, the calculations apply to both $\chi_1$ and $\chi_2$. Since both interact via a Yukawa potential, in the Born limit the cross-section is well posed for $\chi\chibar\to\chi\chibar$. In the center of mass frame, Rutherford-like scattering is obtained \cite{Feng:2009hw,2010PhLB..692...70I}:

\begin{equation}
 \frac{d\sigma}{d\cos\theta} = \frac{\sigma_{0}w^4}{2\lrb{w^2 + v^2\sin^2\lrp{\theta/2}}^2},
\label{eq:rutherford-born-differential-sigma}
\end{equation} 
where $\sigma_0^{i} = 4\pi\alpha_{\chi_i}^2 / \left(m_{\chi,i}^2w^4\right)$ with $\alpha_{\chi_{i}} = (g_{\chi_{i}})^2/4\pi$ and a velocity transition scale, $w = \ma /m_{\chi,i}$, between the constant and power-law fall off of the cross-section; $v$ is the relative velocity between the two initial particles, and $\theta$ is the scattering angle after the particles have interacted. In the case of $\chi\chi\to\chi\chi$, in the center of mass frame, we obtain a M\o{}ller-like differential cross-section~\cite{Yang:2022hkm,Girmohanta:2022dog}:

\begin{equation}
 \frac{d\sigma}{d\cos\theta} = \frac{\sigma_{0}w^4 \left[\left(3\cos^2\theta + 1\right)v^4 + 4v^2w^2 + 4w^4\right]}{\left(\sin^2\theta v^4 + 4v^2w^2 + 4w^4\right)^2}.
\label{eq:moller-born-differential-sigma}
\end{equation} 

\noindent To specify the full model, we have five free parameters: $\lrcb{\mchione,~\mchitwo,~m_{A'},~\gchione,~f}$.


\subsection{Boltzmann Equations}\label{subsec:boltzmann-equations}

\begin{figure}[tbh]
\centering
\includegraphics[scale=0.75]{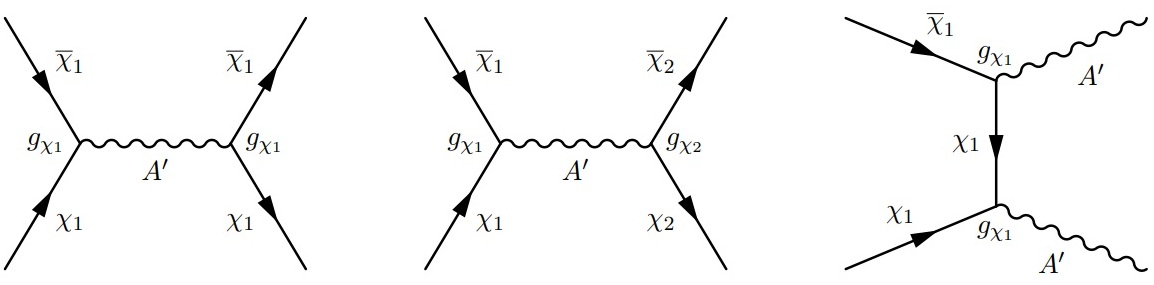}
\vspace{2mm}
\caption{uSIDM $2\to 2$ processes: number conserving (left), number changing (center), $\chibar_1\chi_1/\chibar_2\chi_2\to A'A'$ (right). For the DM to dark photon annihilation, we draw the t-channel; we do not draw the u-channel for simplicity but we do include its contribution to the cross-section. Similarly, for the number conserving $\chibar_1 \chi_1 \to \chibar_1 \chi_1$ scattering, we only show the s-channel diagram, but also the t-channel diagram is included in our analysis.}
\label{fig:usidm-three-panel}
\end{figure}

In order to track the evolution of the dark matter number densities, we can write a Boltzmann equation for each dark matter species, but the key property is that we will need to be careful in how we deal with the annihilation and number changing interaction channels. In particular, we can calculate the branching ratio for the tree level s-channel interaction for $\chi_{1}\chibar_{1}$. From Fig.~\ref{fig:usidm-three-panel} (left), we can see that the invariant scattering amplitude, $\lrv{\M_{11}}^{2}$ is

\begin{equation*}
 \lrv{\M_{11}}^{2} \sim \gchione^{4},
\end{equation*}

\noindent and from Fig.~\ref{fig:usidm-three-panel} (center) we obtain,

\begin{equation*}
 \lrv{\M_{12}}^{2} \sim \gchione^{2}\gchitwo^{2} \sim \gchione^{4}\sqrt{f},
\end{equation*}

\noindent where the last scaling follows from $\gchitwo \sim \gchione f^{1/4}$. So the number changing channel is suppressed by $\sqrt{f}$ compared to the number conserving channel. The corresponding ratio of interaction rates is then:

\begin{equation*}
 \frac{\Gamma\lrp{\chi_{1}\chibar_{1}\to\chi_{2}\chibar_{2}}}{\Gamma_{\rm tot}}\sim \frac{\lrv{\M_{12}}^{2}}{\lrv{\M_{11}}^{2} + \lrv{\M_{12}}^{2}} \sim \frac{\lrv{\M_{11}}^{2}\sqrt{f}}{\lrv{\M_{11}}^{2} + \lrv{\M_{11}}^{2}\sqrt{f}} \sim \frac{\sqrt{f}}{1 + \sqrt{f}} \sim \sqrt{f}~~~\text{for $f \ll 1$},
\end{equation*}

\noindent which is also suppressed by $\sqrt{f}$. This scaling of the amplitudes will aid in solving the Boltzmann equations for $\chi_{1}$ and $\chi_{2}$, which are given by the following system, if we convert to the yield equations, where $Y_{i,eq}(x) \sim x^{3/2}e^{-x}$ for $x = m/T$:

\begin{eqnarray}
 &&\frac{dY_{1}}{dx} = \frac{-s}{Hx}\lrb{\left<\sigma v\right>_{11\to A^{\prime}A^{\prime}}\lrp{Y_{1}^{2} - Y_{1,eq}^{2}} + \left<\sigma v\right>_{11\to22}\lrp{Y_{1}^{2}-\lrp{\frac{Y_{1,eq}^{2}}{Y_{2,eq}^{2}}}Y_{2}^{2}}}\label{eq:chi1_boltzmann_yield},\\
 &&\frac{dY_{2}}{dx} = \frac{-s}{Hx}\lrb{\left<\sigma v\right>_{22\to A^{\prime}A^{\prime}}\lrp{Y_{2}^{2} - Y_{2,eq}^{2}} - \left<\sigma v\right>_{11\to 22}\lrp{Y_{1}^{2}-\lrp{\frac{Y_{1,eq}^{2}}{Y_{2,eq}^{2}}}Y_{2}^{2}}}\label{eq:chi2_boltzmann_yield}.
\end{eqnarray}

In general, one has to solve these coupled equations numerically; however, we can compare the rough order of magnitude of each thermally averaged cross-section. The $\chibar_{1}\chi_{1}\to A^\prime A^\prime$ annihilation will scale as $\gchione^{4}$, whereas the conversion $\bar\chi_{1}\chi_1\to\bar\chi_2\chi_2$ scales as $\gchione^2\gchitwo^2\sim\gchione^4\sqrt{f}$, then $\left<\sigma v\right>_{11\to A^{\prime}A^{\prime}} \gg \left<\sigma v\right>_{11\to 22}$ and we can therefore neglect the second term in the Eq.~\eqref{eq:chi1_boltzmann_yield}. However, for Eq.~\eqref{eq:chi2_boltzmann_yield}, we cannot yet neglect either term, as the $\chibar_{2}\chi_{2}\to A^\prime A^\prime$ annihilation is suppressed by $\sqrt{f}$ compared to the $\bar\chi_{2}\chi_2\to\bar\chi_1\chi_1$ conversion. When we neglect the second term in Eq.~\eqref{eq:chi1_boltzmann_yield}, we obtain a much simpler equation:

\begin{equation}
 \frac{dY_{1}}{dx} = \frac{-s}{Hx}\lrb{\left<\sigma v\right>_{11\to A^{\prime}A^{\prime}}\lrp{Y_{1}^{2} - Y_{1,eq}^{2}}}.
\label{eq:chi1_boltzmann_yield_analytic_ODE}
\end{equation}

\noindent The above equation can be solved analytically to get \cite{Kolb_and_Turner,Gondolo_1991,Ala-Mattinen:2019mpa,Morrison:2020vbz},

\begin{eqnarray}
 &&Y_{1}(\infty)\sim \sqrt{\frac{45}{\pi}}\frac{ x_{f_1}}{\sqrt{g^{*}}M_{\text{pl}}\mchione\left<\sigma v\right>_{11\to A'A'}}\label{eq:chi1_analytic_boltzmann_solution},\\
 &&\Omega_{\chi_1}h^{2} = \frac{h^{2}\mchione s_0 Y_{1}(\infty)}{\rhocrit} \sim 2.742 \times 10^8\left(\frac{\mchione}{\text{GeV}}\right) Y_{1}(\infty),
\label{eq:chi1_analytic_omegah2}
\end{eqnarray}

\noindent where $x_{f_1}$ is the dimensionless ratio of $\mchione / T_{f}$, at the approximate freeze-out temperature, $T_{f}$. That leaves us with the $Y_{2}$ equation. Because of the mass splitting, $r(x)=\lrp{\frac{Y_{1,eq}}{Y_{2,eq}}}^{2} \sim \frac{\mchione^{3}}{\mchitwo^{3}}e^{-2\lrp{\mchione-\mchitwo}/\Tchi} \sim e^{-2\Delta/\Tchi}$, where $\Delta = \mchione - \mchitwo$ is the mass splitting between the two dark matter species. If we take $\mchione > \mchitwo$, then at temperatures $\Tchi \ll \Delta$, the $r(x)Y_{2}^{2}\sim e^{-2\Delta/\Tchi}Y_{2}^{2}$ term is heavily Boltzmann suppressed compared to $Y_{1}^{2}$. Thus, we can write:

\begin{equation}
    \frac{dY_{2}}{dx} = \frac{-s}{Hx}\lrb{\left<\sigma v\right>_{22\to A^{\prime}A^{\prime}}\lrp{Y_{2}^{2} - Y_{2,eq}^{2}} - \left<\sigma v\right>_{11\to 22}Y_{1}^{2}}.
\label{eq:chi2_yield_suppressed}
\end{equation}

This holds for when $r(x)Y_{2}^{2} \ll Y_{1}^{2} \implies r(x) \ll \lrp{\frac{Y_{1}}{Y_{2}}}^{2}$, but since $\fend \sim \frac{Y_{1}}{Y_{1} + Y_{2}}\sim\frac{Y_{1}}{Y_{2}}$, then the condition on $r(x)$ becomes $r(x)\ll \lrp{\fend}^{2}$, which gives a condition on the mass splitting, $\Delta$: $e^{-2\Delta/\Tchi} \ll \lrp{\fend}^{2}$, giving

\begin{equation}
    \frac{\Delta}{\Tchi} \gtrsim \ln{\lrp{\frac{1}{\fend}}}.
\label{eq:general_mass_split_condition}
\end{equation}

We can estimate this ratio by taking $\fend \sim 10^{-3.5}$ which is motivated to explain high-z quasars and LRDs~\cite{Roberts_LRD}. Such a value implies $\frac{\Delta}{\Tchi} \gtrsim 8$. Taking $\Tchi \sim T_{f_2} = \mchitwo/x_{f_2}$, where $x_{f_2}$ is the dark sector freeze out value, then the mass splitting can be written as:

\begin{equation}
    \Delta \gtrsim 8\frac{\mchitwo}{x_{f_2}}\sim 0.62\lrp{\frac{13}{x_{f_2}}}~\mchitwo.
\label{eq:delta_condition}
\end{equation}

However, Eq.~\eqref{eq:chi2_yield_suppressed} still contains a term proportional to the conversion cross-section, $\left<\sigma v\right>_{11\to22}Y_{1}^{2}$. For freeze out, we are interested in the solution as $x\to\infty$, so the quantity that matters is whether or not $\left<\sigma v\right>_{11\to22}Y_{1}^{2} \ll \left<\sigma v\right>_{22\to A'A'}Y_{2}^{2}$. This implies that we need to look at the ratio:

\begin{eqnarray*}
    &&\frac{\left<\sigma v\right>_{11\to22}}{\left<\sigma v\right>_{22\to A'A'}}\frac{Y_{1}(\infty)^{2}}{Y_{2}(\infty)^{2}} \sim \frac{1}{f^{1/2}}\lrp{\fend}^{2} \ll 1, \quad \Rightarrow \quad \fend \ll f^{1/4} ~, 
\end{eqnarray*}
where as before, the ratio of the yields gives the factor of $\lrp{\fend}^{2}$, and the cross-section ratio gives the factor $\frac{\gchione^{4}f^{1/2}}{\gchitwo^{4}} = \frac{f^{1/2}}{f} = \frac{1}{f^{1/2}}$. 

Given that uSIDM requires $\fend\sim10^{-3.5}$ to explain high-z quasars and LRDs, the above conditions hold so long as $f \gg \lrp{\fend}^{4} \sim (10^{-3.5})^{4} \sim 10^{-14}$, which is a very weak bound. Therefore, for all reasonable values of $f$, the above conditions hold, and we can drop the conversion term from the $Y_{2}$ equation:

\begin{equation}
    \frac{dY_{2}}{dx} = \frac{-s}{Hx}\lrb{\left<\sigma v\right>_{22\to A^{\prime}A^{\prime}}\lrp{Y_{2}^{2} - Y_{2,eq}^{2}}},
\label{eq:chi2_boltzmann_yield_final}
\end{equation}

\noindent which can be solved in the usual way via Eq.~\eqref{eq:chi1_analytic_boltzmann_solution} and Eq.~\eqref{eq:chi1_analytic_omegah2}. The total energy density in the dark sector is then given by:

\begin{equation}
 \OmegaDMh = \Omega_{\chi_1}h^{2} + \Omega_{\chi_2}h^{2}.
\label{eq:total_omegaDM}
\end{equation}

\noindent We can also calculate the freeze-out value of the uSIDM fraction, i.e., the amount of the total dark matter that is left over after freeze-out to collapse into SMBH seeds:

\begin{equation}
 \fend = \frac{\Omega_{\chi_1}h^{2}}{\Omega_{\chi_1}h^{2} + \Omega_{\chi_2}h^{2}}.
\label{eq:final-uSIDM-fraction}
\end{equation}

\subsection{Thermally Averaged Cross-Sections}\label{sec:thermally-averaged-cross-sections}

In order to solve Eq.~\eqref{eq:chi1_analytic_boltzmann_solution} and Eq.~\eqref{eq:chi2_boltzmann_yield_final}, we need to know the thermally averaged cross-section for each process \cite{Gondolo_1991}:

\begin{equation}
 \left<\sigma v\right> = \frac{1}{8\mchi^{4}TK_{2}^{2}\lrp{\mchi/T}}\int_{4\mchi^{2}}^{\infty}\sigma(s)\lrp{s-4\mchi^{2}}\sqrt{s}K_{1}\lrp{\frac{\sqrt{s}}{T}}ds,
\label{eq:gondolo_thermal_average}
\end{equation}

\noindent which means that we need to calculate the center of mass cross-section for each process in Fig.~\ref{fig:usidm-three-panel} (center) and Fig.~\ref{fig:usidm-three-panel} (right). Given that the full thermal averaging is numerically expensive to compute, we follow Ref.~\cite{Cannoni_sigma_vrel} and expand $\left<\sigma v\right>$ in inverse powers of $x$ for all channels. In general, for species which are non-relativistic during freeze-out, as our particles are, one need only keep terms up to $O(x^{-1})$ in the expansion:

\begin{equation}
 \left<\sigma v\right> \sim a + \frac{6b}{x} + \cdots.
\end{equation}

In doing so, the annihilation cross-section (for both $\chi_1$ and $\chi_2$) can be expanded into both s-wave and p-wave pieces to good approximation. However, in our case, the s-wave term will be dominant over the p-wave term; thus, for the annihilation cross-section, the expansion in velocity is simply given by ($w = \ma/\mchi$):

\begin{eqnarray}
&&\sigma v_{\rm rel}~=~ a(w) + \mathcal O(v_{\rm rel}^2),\\
&&a(w)=\frac{\pi \alphachi^2}{m_\chi^2}\,\frac{(1-w^2)^{3/2}}{(1-\tfrac{w^2}{2})^2}~.
\end{eqnarray}

\noindent In the non-relativistic limit, $s=4m_\chi^2(1+\tfrac{v^2}{4}+\cdots)$, Eq.~\eqref{eq:gondolo_thermal_average} reduces to the Maxwell-Boltzmann thermal average,

\begin{equation}
\big\langle \sigma v \big\rangle \simeq
\frac{x^{3/2}}{2\sqrt{\pi}}
\int_0^\infty v^2 e^{-x v^2/4}~(\sigma v)~dv ,
\label{eq:maxwell-boltzmann-avereage}
\end{equation}

\noindent which allows us to take the velocity expansion and convert to an expansion in the parameter $x = m / T$. Thus, the non-relativistic thermally averaged cross-section is given by:

\begin{equation}
 \left<\sigma v\right>(x) = a(w), 
\end{equation}

\noindent with $a(w)$ given above. With the above expansion, we can solve Eq.~\eqref{eq:total_omegaDM} analytically, under the assumption that $w \ll 1$, for the relic abundance relation between $\mchitwo$ and $\alphachitwo$. From the SIDM literature \cite{Tulin:2013teo,Adhikari:2022sbh} and references therein, the typical SIDM thermal relic used is $\lrp{\mchi/\rm GeV} \sim \alphachi / \lrp{4\times10^{-5}}$. We should expect some modulation to this relation in our uSIDM picture, but in the limit that $f\to0$, they should be consistent. By using the form of Eq.~\eqref{eq:chi1_analytic_boltzmann_solution} and Eq.~\eqref{eq:chi1_analytic_omegah2} for both $Y_{1}$ and $Y_{2}$ into Eq.~\eqref{eq:total_omegaDM}, and using that $a(w)\to \frac{\pi\alphachi^2}{\mchi^2}$ for $w \ll 1$, one finds:

\begin{equation*}
 \OmegaDMh = \frac{1.07\times10^9}{\sqrt{g^{*}}M_{\text{pl}} \pi} \lrb{x_{f_1}\frac{\mchione^2}{\alphachione^2} + x_{f_2}\frac{\mchitwo^2}{\alphachitwo^2}}.
\end{equation*}

\noindent Using $\gchitwo \sim \gchione f^{1/4}$, we can simplify the ratio $\lrp{\alphachitwo / \alphachione}^2 = f$, and using the mass splitting criterion, Eq.~\eqref{eq:general_mass_split_condition}, we then solve for $\mchitwo$ to get:

\begin{equation*}
 \mchitwo = \alphachitwo \sqrt{\frac{\OmegaDMh}{\frac{1.07\times10^{9}}{\sqrt{g_\star}\mpl\pi}}}\frac{1}{\sqrt{x_{f1} f (1 - \frac{1}{x_{f_2}}\ln\lrp{\fend})^{2}+ x_{f2}}}.
\end{equation*}

\noindent With $\OmegaDMh = 0.12$, $g^{*} \sim 60$, and $M_{\text{pl}} = 1.22\times10^{19}~\rm GeV$, the above equation reduces to,

\begin{equation}
 \frac{\mchitwo}{\rm GeV} \sim \frac{\alphachitwo}{1.8\times10^{-5}}\sqrt{\frac{13}{x_{f_2}}}\lrp{1 -\delta},\\
\label{eq:uSIDM_relic_density}
\end{equation}

\noindent where $\delta\sim4\times10^{-7}\lrp{\frac{f}{10^{-4.1}}}\lrp{\frac{x_{f_1}}{22}}\lrp{\frac{13}{x_{f_2}}}^{3}\lrb{\frac{1}{13}\lrp{\frac{13}{x_{f_2}}}-8\lrp{\frac{\ln\lrp{\fend}}{8}}}^{2}$. Thus, as $\delta\ll1$, the uSIDM particle only introduces a small correction onto the usual SIDM thermal relic relation. Similarly, we can calculate an analytic expression for the final uSIDM fraction via Eq.~\eqref{eq:final-uSIDM-fraction}:

\begin{equation}
 \fend = \frac{x_{f_1}\frac{\mchione^2}{\alphachione^2}}{x_{f_1}\frac{\mchione^2}{\alphachione^2} + x_{f_2}\frac{\mchitwo^2}{\alphachitwo^2}} \sim f\frac{x_{f_1}}{x_{f_2}}\lrp{1 - \frac{1}{x_{f_2}}\ln\fend}^{2}~.
\label{eq:f_uSDIM_end}
\end{equation}

From Ref.~\cite{Roberts_LRD}, in order to correctly account for the number density of Little Red Dots, $\fend\sim10^{-3.5}$; this implies a value of $f\sim10^{-4.1}$. We use these values as fiducial estimates throughout this work.

\section{Results}\label{sec:results}

\subsection{Thermal Relic Density Abundance}\label{subsec:thermal-relic-density-abundance}

\begin{figure}[tbh] 
\centering
\includegraphics[scale=0.5]{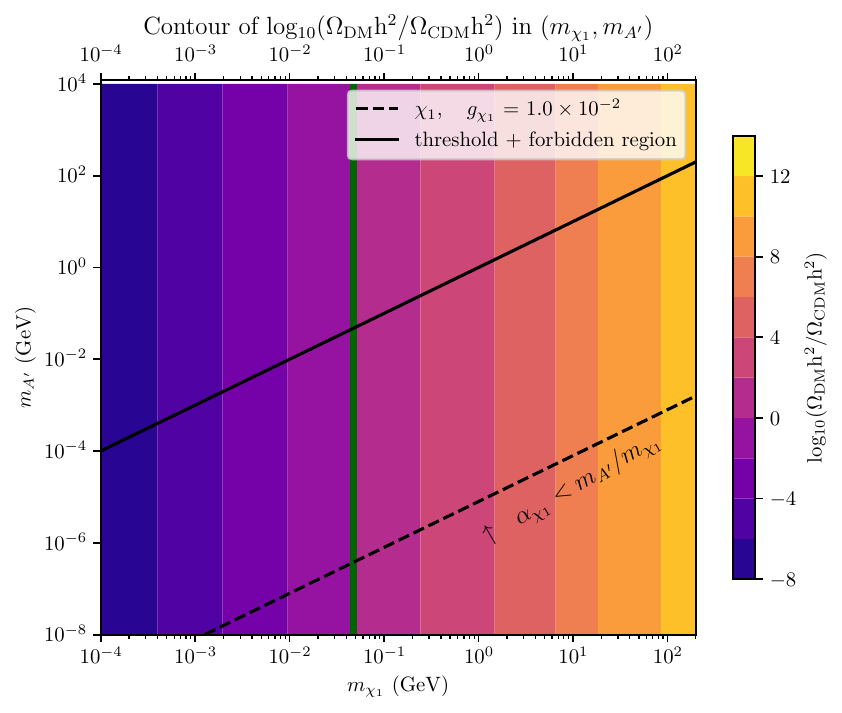}
\includegraphics[scale=0.5]{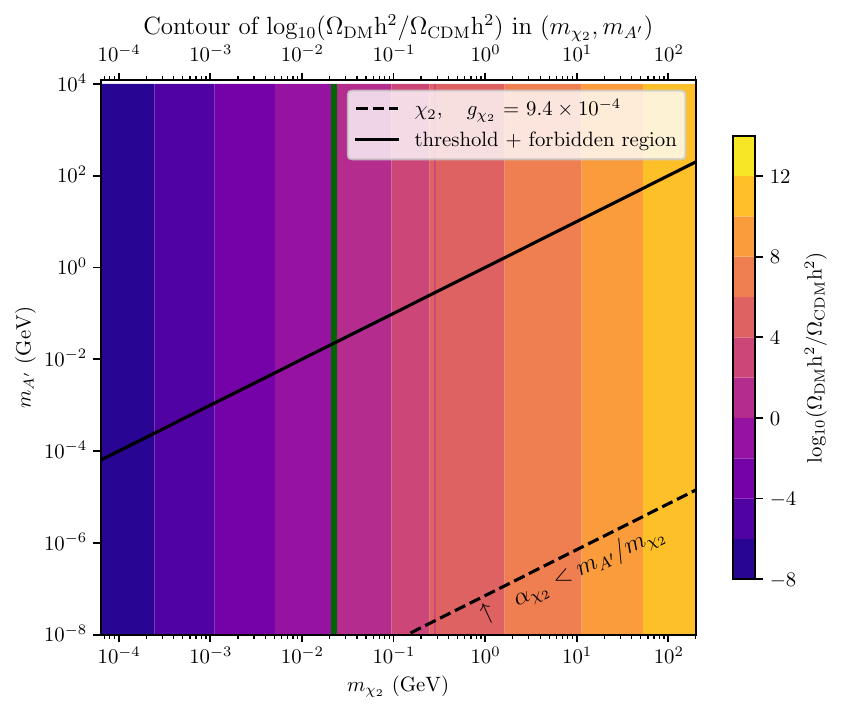}
\caption{The $\ma$ - $\mchione$ parameter plane (left) and $\ma$ - $\mchitwo$ (right) for fixed coupling constants $\gchione$ and $\gchitwo$. The filled level contours show the dark matter relic abundance $\OmegaDMh$. Above the black dashed line in each plot, the $\chi_1$ or $\chi_2$ is in the Born region. The green line corresponds to $\rm \log_{10}\left(\OmegaDMh/\OmegaCDMh\right) = 0$, i.e., the model yields the correct observed dark matter relic density. The black solid line shows the threshold for the $\chi\chibar \to A^\prime A^\prime$ annihilation to occur. Note that because of thermal energy, for smaller $\mchi$ the production can occur above threshold; however, we expect our approximations to break down in this region.}
\label{fig:parameter-space-relic-abundance}
\end{figure}

To get an understanding of the viable parameter space of the uSIDM model, as well as see the various kinematic thresholds, we scan over a grid of $\lrp{\mchione,~\ma}$ using Eq.~\eqref{eq:total_omegaDM} and set the $\mchione$ and $\mchitwo$ relation via Eq.~\eqref{eq:delta_condition}. In Fig.~\ref{fig:parameter-space-relic-abundance}, we plot contours of $\rm{log_{10}\lrp{\OmegaDMh/\Omega h^2}}$ for fixed coupling constants $\gchione$ and $\gchitwo$ corresponding to $f = 10^{-4.1}$. The green band overlays the contour corresponding to $\OmegaDMh = 0.12$. For convenience we plot the Born region ($\alphachi < \ma/\mchi$) dividing cutoff for $\chi_1$ and $\chi_2$ as dashed black lines respectively. We find that both species of particles are almost always in the Born region. However, as we derived in Eq.~\eqref{eq:uSIDM_relic_density}, this is the relation that drives the actual constraint on the relic density, rather than simply scanning over values of $\lrp{\mchione,~\ma}$ with a fixed $\alphachione$ and $\alphachitwo$.

\subsection{SIDM Cross-section Constraints}\label{subsec:SIDM-cross-section-constraints}

\begin{figure}[tb]
\includegraphics[scale=0.65]{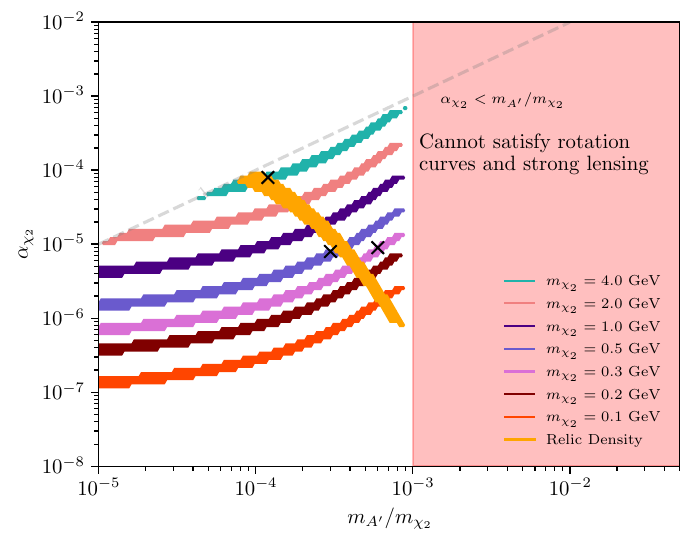}
\includegraphics[scale=0.65]{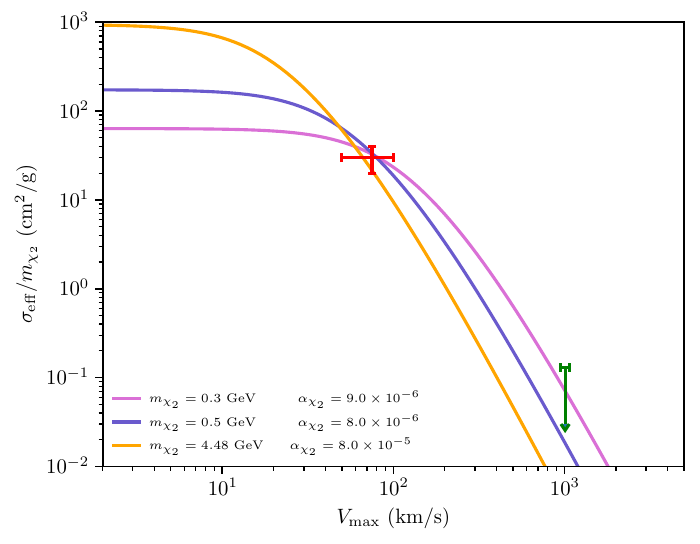}
\caption{{\bf Left Panel:} The favored SIDM parameter space for considerations of the relic density constraint Eq.~\eqref{eq:uSIDM_relic_density} (orange contour) and fixed $\mchitwo$ values, specified by the corresponding colors shown in the legend. The gray line corresponds to the transition between the quantum region and the Born region, i.e., below the gray line - the weakly-coupled Born approximation is valid, whereas above one must solve the Schrodinger equation for the Yukawa potential via partial wave analysis \cite{Tulin:2013teo}. The contours correspond to satisfying the dwarf/low-surface brightness galaxy rotation curve $\sigmaeffmchitwo=20\textup{--}40~{\cmg}$ at $\vmax = (75\pm 25)~\kms$~\cite{Roberts_rotation_curve}, and strong cluster lensing upper bounds, represented by $\sigmaeffmchitwo < 0.13~\cmg$ at $\vmax=(1000\pm 55)~\kms$ \cite{Andrade:2020lqq}. {\bf Right panel:} The effective cross-section as a function of $\vmax$ for the representative points in the figure legend that correspond to the black crosses in the left panel.}
\label{fig:roberts2025_rotation_curve_plots}
\end{figure}

Thus, to more realistically constrain the viable parameter space, we replicate contours of SIDM cross-section constraints from rotation curves of dwarf and low-surface brightness galaxies (LSBs), as well as strong lensing in galaxy clusters from Ref.~\cite{Roberts_rotation_curve} in Fig.~\ref{fig:roberts2025_rotation_curve_plots} (left). Overplotted in the orange contour is our derived modified relic density relation (Eq.~\eqref{eq:uSIDM_relic_density}). In Fig.~\ref{fig:roberts2025_rotation_curve_plots} (right), we plot three representative cases (the black crosses in the left panel), to showcase the diversity of effective cross-section versus the maximum circular velocity, $\vmax$, curves in the SIDM parameter space ~\cite{Outmezguine:2022bhq,Yang:2022hkm,Yang:2022zkd,Fischer:2023lvl}. The effective cross-section is defined as,
{\everymath{\displaystyle}
\begin{eqnarray}
 \sigmaeff &=& \frac{2\int v^2 dv~d\cos\theta \frac{d\sigma}{d\cos\theta}v^5\sin^{2}\theta ~\text{exp}\left[-\frac{v^2}{4\left(\sigmavel\right)^2}\right]}{\int v^2 dv~d\cos\theta v^5\sin^{2}\theta ~\text{exp}\left[-\frac{v^2}{4\left(\sigmavel\right)^2}\right]}  \nonumber\\
 &\approx& \frac{1}{512\left(\sigmavel\right)^8}\int v^2 dv~\frac{2}{3}\sigma_{V}v^5 ~\text{exp}\left[-\frac{v^2}{4\left(\sigmavel\right)^2}\right],
\label{eq:sigmaeff}
\end{eqnarray}}
where $\sigmavel \approx0.64 \vmax$ is the characteristic velocity dispersion of the halo. The viscosity cross-section, $\sigma_{V}$, is given by
\begin{equation}
 \sigma_{V} = \frac{3}{2}\int d\cos\theta \sin^{2}\theta \frac{d\sigma}{d\cos\theta}.
\label{eq:definition-viscosity-cross-section}
\end{equation}

\noindent As we are considering scattering, we model the differential cross-section as a Rutherford process via Eq.~\eqref{eq:rutherford-born-differential-sigma}. Doing the angular integration for $\sigma_{V}$ one finds:

\begin{equation}
 \sigma_{V} = \frac{6\sigma_{0}w^{6}}{v^{6}}\lrb{\lrp{2 + \frac{v^2}{w^2}}\ln\lrp{1 + \frac{v^2}{w^2}} - \frac{2v^2}{w^2}}
\label{eq:final-viscosity-cross-section}.
\end{equation}

M\o{}ller scattering is also present; however, as a simplifying approximation, we only consider Rutherford scattering. We do not expect the results to change qualitatively by including both channels. For the chosen relic density case, shown by the orange curve in Fig.~\ref{fig:roberts2025_rotation_curve_plots} (right), the model is able to access very high cross-sections at low $\vmax$. Therefore, the overall model is able to be consistent with both early universe and late universe SIDM constraints.

\subsection{Linear Matter Power Spectrum}\label{subsec:matter-power-spectrum}

\begin{figure}[tb]
\centering
\includegraphics[scale=0.7]{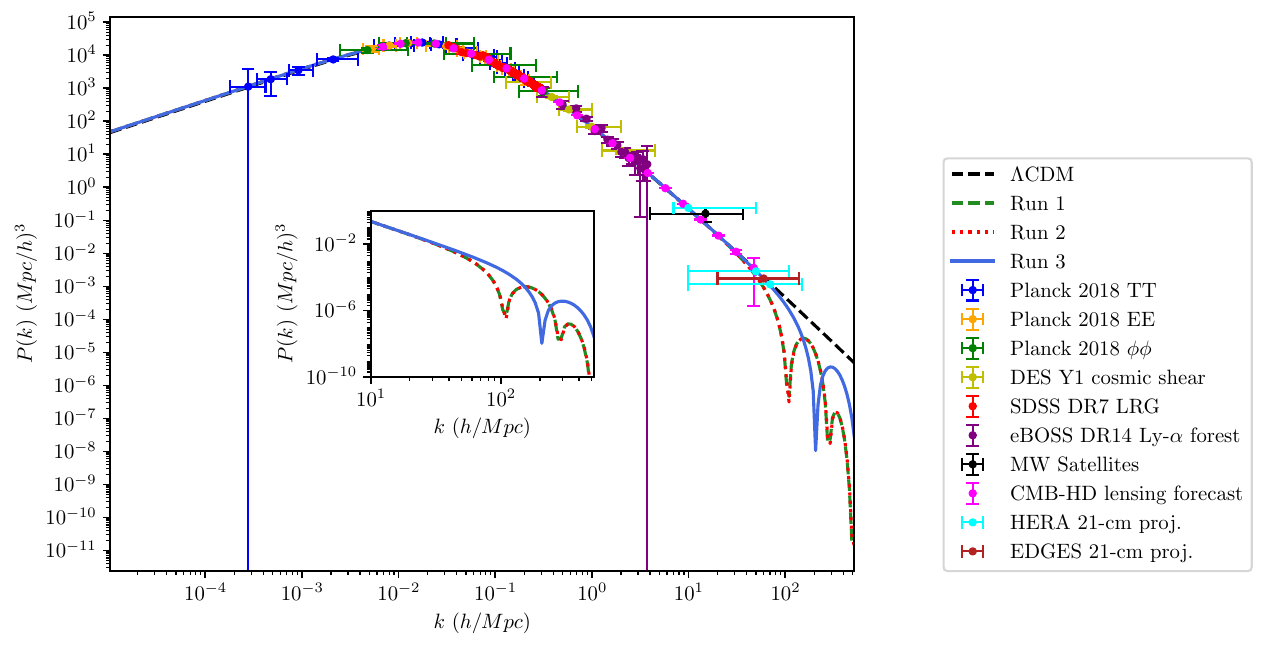} \\[8pt]
\includegraphics[scale=0.7]{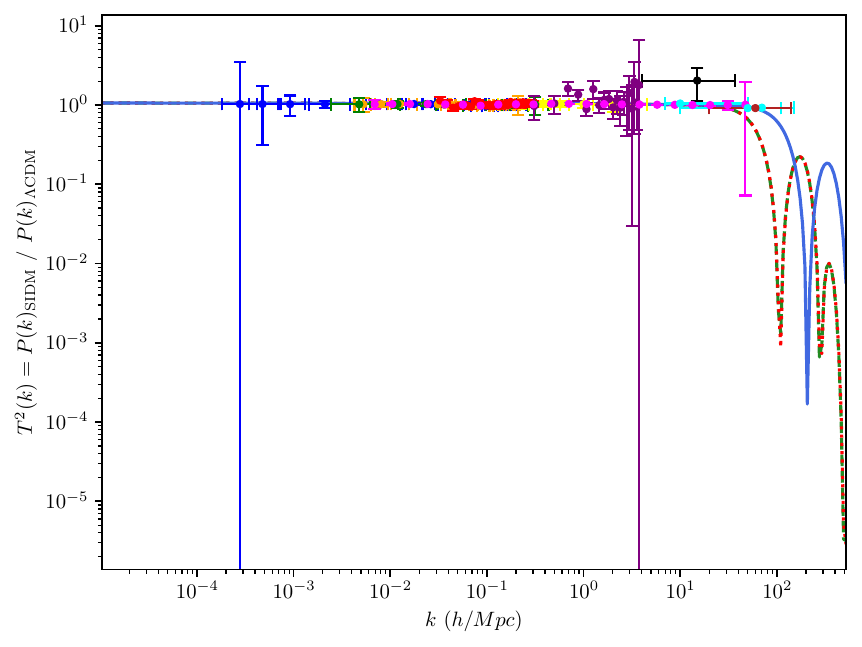}
\caption{Using the three chosen parameter sets from Table~\ref{tab:power-spectrum-parameters} (dashed green, dotted red, solid blue), we plot the corresponding matter power spectrum (top) and transfer function (bottom). The data are from Planck 2018~\cite{Planck:2018nkj,Chabanier:2019eai}, DES Y1~\cite{DES_Y1}, SDSS DR7~\cite{SDSS_DR7}, eBOSS DR14~\cite{Abolfathi_2018}, MW Satellites~\cite{MW_forecast}, and future projected measurements via CMB-HD lensing~\cite{MacInnis_2025}, and the HERA and EDGES 21-cm projections~\cite{HERA,EDGES,Munoz:2019hjh}.}
\label{fig:matter-power-spectrum}
\end{figure}

\begin{table}[tb]
 \centering
\begin{tabular}{|c|c|c|c|c|c|c|}
\hline
Run & Curve Style & $\mchitwo$ [GeV] & $\ma$ [GeV] & $\alphachitwo$ & $\alpha_{\psi}$ & $\alphaethos$ [$\text{Mpc}^{-1}$]\\\hline
1 & Dashed Green & $0.3$ & $1.8\times10^{-4}$ & $9\times10^{-6}$ & $9\times10^{-6}$ & $\approx85.1$\\
2 & Dotted Red & $0.5$ & $1.5\times10^{-4}$ & $8\times10^{-6}$ & $8\times10^{-6}$ & $\approx83.7$ \\
3 & Solid Blue & $4.48$ & $5.4\times10^{-4}$ & $8\times10^{-5}$ & $8\times10^{-5}$ & $\approx5.7$ \\
\hline
\end{tabular}
\caption{Parameters used to generate the linear matter power spectrum shown in Fig.~\ref{fig:matter-power-spectrum}.}
\label{tab:power-spectrum-parameters}
\end{table}

Generically, one expects SIDM models to cause a suppression of the linear matter power spectrum at a certain critical wave number, but also for dark acoustic oscillations (DAOs), in analogy to the baryon acoustic oscillations \cite{Tulin:2017ara,Adhikari:2022sbh}. To explore this possibility, we allow the dark photon to decay to a massless fermion species, $\psi$, which acts as the dark radiation species via,

\begin{equation*}
    \lag_{\rm{int}} = -g_{\psi}\bar{\psi}A'^{\mu}\gamma_{\mu}\psi~.
\end{equation*}

Following Ref.~\cite{Huo_2018}, for simplicity, we take $g_{\psi} = \gchitwo$, which causes the suppression and the resulting DAOs to appear at much higher wave numbers than are currently probed by existing experiments, with a small subregion constrained by MW Satellite measurements. Future projected experiments would begin to increasingly measure the non-standard features predicted in our model~\cite{MacInnis_2025,HERA,EDGES,Munoz:2019hjh}. Given that our SIDM species couples more weakly to the dark photon than the uSIDM species, which then decays to the dark radiation species, it is possible to enhance the suppression scale to be closer to the current and upcoming measurements of the power spectrum. The decay of the dark photon to the dark radiation species is rapid and hence avoids overclosing the universe with a massive dark photon. The decay width for a vector decaying to massless Dirac fermions is, $\Gamma_{A'\lra\bar{\psi}\psi} = \frac{1}{3}\alpha_{\psi}\ma$, which for our fiducial parameters given in Table~\ref{tab:power-spectrum-parameters}, corresponds to lifetimes $\tau_{A'\lra\bar{\psi}\psi} \approx 10^{-17}-10^{-15}~\text{s}$.

We use \texttt{CLASS} \cite{CLASSII,CLASS_with_ETHOS} to calculate the linear matter power spectrum from the SIDM particle. To interface with \texttt{CLASS}, we use the \texttt{ETHOS} \cite{CLASS_with_ETHOS} parameterization input to \texttt{CLASS}. This means that we must specify the dark matter particle mass, and \texttt{ETHOS} effective coupling, $\alphaethos$. From Ref.~\cite{CLASS_with_ETHOS}, we can write

\begin{equation}
 \alphaethos = 3.3~\text{Mpc}^{-1}~\lrp{\frac{10^{-4}~\rm{GeV}}{\ma}}^{4}\lrp{\frac{1~\rm{GeV}}{\mchitwo}}\lrp{\frac{\alphachitwo}{10^{-6}}}\lrp{\frac{\alpha_{\psi}}{10^{-6}}}\lrp{\frac{\xi}{0.3}}^{2},
\end{equation}

\noindent where $\xi = T_{\chi} / T_{\rm{CMB}}$ is the ratio of the dark sector temperature to the CMB temperature. In Fig.~\ref{fig:matter-power-spectrum} we plot the corresponding linear matter power spectrum (top) and transfer function (bottom), for the parameters given in Table~\ref{tab:power-spectrum-parameters}; we also set $\xi = 0.3$ as our fiducial value in each case, which keeps $\Delta N_{\rm{eff}}$ within current bounds of $\Delta N_{\rm{eff}} \lesssim 0.107$ \cite{best_Delta_Neff_constraint}. With increasing $\xi$, the DAOs will occur at a lower wave number, whereas if we decrease $\xi$,  the DAOs will occur at a larger wave number. We chose these parameters as they correspond to the shown example orange curve from the thermal relic abundance constraint in Fig.~\ref{fig:roberts2025_rotation_curve_plots} (right), as well as other parameters within the relic abundance contour in Fig.~\ref{fig:roberts2025_rotation_curve_plots} (left). To understand the impact on the power spectrum from the uSIDM particle, we run \texttt{CLASS} with a dark matter interacting fraction $\fend\sim10^{-3.5}$. Because the fraction of particles that interact is so small, there is negligible impact on the power spectrum from the uSIDM particles until much larger wavenumbers (smaller scales), $k \gtrsim 300~(\rm{h/Mpc})$. From this analysis we have verified that the uSIDM driven DAOs will happen at a significantly larger wave number compared to the SIDM driven DAOs.


\section{Conclusion}\label{sec:conclusion}

We have presented a minimal two--component self-interacting dark matter framework in which an ultra-strongly self-interacting subcomponent (uSIDM) coexists with a dominant, moderately self-interacting SIDM component. The interactions are mediated by a light dark photon, and the model is specified by the parameter set $\{\mchione, \mchitwo,m_{A'},g_{\chi_1},f\}$,  where $f = \gchitwo^4/\gchione^4$ denotes the ratio of the couplings of the two dark matter species to the fourth power and corresponds approximately to the abundance fraction of uSIDM. The key phenomenological inputs are (1) the uSIDM coupling constant, $g_{\chi_1}$, and (2) the mass ratio between the dark photon and dark matter species, $\ma/\mchione$ and $\ma/\mchitwo$.

On the particle-physics side, we derived the relevant thermally averaged cross-sections and solved the coupled Boltzmann system for the two species to determine the relic abundance. Using an explicit non-relativistic expansion, we showed that the conversion channel $\bar \chi_2\chi_2\to \bar\chi_1\chi_1$ is highly suppressed due to the mass splitting $m_{\chi_1} > m_{\chi_2}$ in the parameter space of interest. This justifies the neglect of this channel in the freeze-out dynamics for our benchmarks. In turn, this leads to simple analytic expressions for the relic density which establish a clear relation between $\alpha_{\chi_2}$ and $\mchitwo$ in  Eq.~\eqref{eq:uSIDM_relic_density} as well as a compact expression for the late-time uSIDM fraction in Eq.~\eqref{eq:f_uSDIM_end}, up to the small corrections.

On the astrophysics side, we confronted the model with a set of complementary probes. We computed the viscosity cross-section $\sigma_V$ for Yukawa-potential scattering and evaluated an effective cross-section $\sigma_{\rm eff}(V_{\max})$ using the velocity distribution appropriate for halos. Imposing $\sigmaeffmchitwo = 20-40~\cmg$ at $\vmax=75\pm25~\kms$ while requiring $\sigmaeffmchitwo<0.13~\cmg$ at $\vmax=1000\pm55~\kms$, we identified viable regions in $(\alphachitwo,~\ma/\mchitwo)$ parameter space consistent with dwarf and cluster-scale observations. We computed the transfer function and verified that the scales affected by the new relativistic degrees of freedom and subpercent uSIDM fraction remain compatible with large-scale structure and Lyman-$\alpha$ considerations for our benchmarks. As the uSIDM is depleted into black-hole seeds at early times, only the SIDM component of DM contributes today.

The resulting overlays imply a mediator-to-DM mass ratio requirement and a corresponding range for the dark matter interaction strength for viable parameter space,

\begin{equation}
10^{-4} \lesssim \frac{\ma}{\mchitwo}\lesssim 10^{-3}~, \quad  10^{-6} \lesssim \alphachitwo \lesssim 10^{-4}~,
\end{equation}
for which: (1) the relic density of the dominant SIDM is obtained with moderate couplings as per Eq.~\eqref{eq:uSIDM_relic_density}, (2) the effective self-scattering satisfies requirements at dwarf galaxy and galaxy cluster scales, and (3) the linear matter power spectrum remains within current bounds. Within this window, the subpercent uSIDM fraction can achieve the ultra-strongly self-scattering required for rapid gravothermal evolution in early halos, providing a path to early black-hole seeding without disrupting late-time small-scale structure traced by the dominant SIDM.

Several directions follow directly from the ingredients developed here. A more exhaustive mapping of $\sigma_{\rm eff}(V_{\max})$ to rotation-curve diversity using the same microphysical inputs would provide further insights into the viable parameter space of the uSIDM model. Additionally, more advanced modeling of cluster members using stellar kinematics as well as more precise observations of strong-lensing in galaxy clusters will further clarify the SIDM impacts on strong-lensing which can lead to further constraints in the uSIDM parameter space. The calculations performed here on the impacts of uSIDM on the linear matter power spectrum can be expanded to include a wider range of parameter space in $f$ and $\xi$. Each of these can be addressed without extending the particle content, using the formalism we have established.

The minimal two-component uSIDM model presented here admits a simple, analytic model of the relic density and a clear organizing role for the mass ratios $\ma/\mchione$ and $\ma/\mchitwo$. The phenomenology of uSIDM is simultaneously consistent with dwarf/LSB galaxy and cluster observations, linear matter power spectrum constraints, and $\Delta N_{\rm{eff}}$ bounds. Coupled with the ability to produce Little Red Dots in the correct abundance, we have shown that the uSIDM model is both cosmologically consistent as well as phenomenologically compelling.


\section{Acknowledgments}

We are grateful to Pankaj Munbodh for helpful comments and references and to Stefania Gori and Pouya Asadi for helpful discussion, feedback, and comments. This work is supported, in part, by the U.S.\ Department of Energy grant number de-sc0010107.


\appendix 

\section{Bound State Estimate}\label{appendix:bound-state-estimate}

Due to the interaction between the two dark matter species, one could imagine that they form bound states. This could either be in the form of unstable $\chi_1 \bar \chi_1$ and $\chi_2 \bar \chi_2$ bound states, or in the form of stable ``dark atoms'' of the uSIDM species $\chi_1$ with one or more SIDM particles $\chi_2$, such that the effective charge of $\chi_1$ gets partially or completely screened. Here we show that this is not the case. The scattering rate for bound state formation (BSF) can be written as $\Gamma_{\rm BSF} = n_{\chi}\langle\sigma_{\rm BSF} v\rangle$. We estimate the number density as:

\begin{equation}
 n_{\chi} \sim \frac{\rho_{\chi}}{\mchi}\lrp{\frac{T_{\chi}}{\xi T_{\rm{CMB}}}}^{3}~,
\end{equation}

\noindent where $\xi = T_{\chi}/T_{\rm{CMB}}$; $T_{\chi}$ and $T_{\rm{CMB}}$ are the dark sector temperature and the CMB temperature today respectively. The BSF cross-section can be parameterized as \cite{Cyr-Racine:2012tfp,vonHarling:2014kha}:

\begin{equation}
 \langle\sigma_{\rm BSF} v\rangle \sim \lrp{\frac{\alphachi}{\mchi}}^{2}\lrp{\frac{B_{\chi}}{T_{\chi}}}^{1.5}
\end{equation}

\noindent where $B_{\chi}$ is the dark binding energy of the bound state. We can estimate $B_{\chi} \sim \frac{1}{2}\alphachi^{2}\mchi$. The Hubble expansion rate as a function of temperature we can write as:

\begin{eqnarray}
 H(T) = \sqrt{\frac{8 \pi^{3} g_{*}(T)}{90}}\frac{T^{2}}{M_{\rm pl}}.
\end{eqnarray}

We now evaluate the above equations under the condition that $T_{\chi} = B_{\chi}$. If $\Gamma_{\rm BSF} > H(T)$, then the formation of bound states is efficient and we will necessarily form ``dark atoms''. If $\Gamma_{\rm BSF} < H(T)$, then the expansion rate of the universe is too large for the particles to form bound states. Applying the above equations, we find a condition on $\alphachi$ such that $\Gamma_{\rm BSF} < H(T)$:

\begin{equation}
 \alphachi^4 < 2\sqrt{\frac{8 \pi^{3} g_{*}(B_{\chi}/\xi)}{90}}\frac{\mchi^2\xi T_{\rm{CMB}}^3}{M_{\rm pl}\rho_{\chi}}.
\end{equation}

Using some fiducial values, $T_{\rm{CMB}} = 2.35\times10^{-13}~\rm{GeV}$, $M_{\rm pl} = 1.22\times10^{19}~\rm{GeV}$, $\rho_{\chi} = 4\times10^{-47}~\rm{GeV}^4$, $\xi = 0.3$, and $\mchi\sim1~\rm{GeV}$, gives $\alphachitwo \lesssim 3\times10^{-3}$. As we have constrained $\alphachitwo$ to be in the range of $10^{-6}$ to $10^{-4}$, the SIDM particle will not form any $\chi_2 \bar \chi_2$ bound states. As for the uSIDM particle, using the same fiducial values (but with a reduced density, $\rho_{\chi_1} \approx \fend\rho_{\chi}$) gives $\alphachione \lesssim 2.3\times10^{-2}$. Thus, no uSIDM $\chi_1 \bar \chi_1$ bound states are formed either. Using the values from Table~\ref{tab:power-spectrum-parameters}, we also find that $\Gamma_{\rm BSF} < H(T)$. Analogously, there will also be no stable $\chi_1 \bar \chi_2$ or $\bar \chi_1 \chi_2$ bound states in the region of parameter space we are interested in.


\bibliographystyle{JHEP}

\bibliography{references.bib}
\end{document}